# Polarimetric Study of GRS 1915+105: Estimation of Interstellar Polarization Component

Jin Inokuchi [1], Koji S. Kawabata [2,3], Makoto Uemura [2,3], Hiroyuki Hiraga [1]

1. Hiroshima University High School, Fukuyama, 5-14-1 Kasuga-cho, Fukuyama, Hiroshima 721-8551, Japan

2. Hiroshima Astrophysical Science Center, Hiroshima University, 1-3-1 Kagamiyama, Higashi-Hiroshima, Hiroshima 739-8526, Japan
3. Physics Program, Graduate School of Advanced Science and Engineering, Hiroshima University, 1-3-1 Kagamiyama, Higashi-Hiroshima, Hiroshima 739-8526, Japan

**Abstract**
GRS 1915+105 is a well-known X-ray binary system composed of a black hole with a low-mass companion star and is recognized for emitting relativistic jets. Imazato et al. (2021) performed extensive polarimetry in a near-infrared (NIR) $K_s$ band from 2019 April through December when GRS 1915+105 experienced an X-ray low luminous state and found almost stable polarization of $P = 2.42\% \pm 0.08\%$. We performed NIR polarimetry of the field stars around GRS 1915+105 in 2023 April and October, and found that the field stars that are not listed in Gaia DR3 and StarHorse2 catalogues show well aligned polarization that is consistent with GRS 1915+105's polarization. Those suggest that the interstellar clouds existing beyond 4 kpc causes the large interstellar extinction and that the polarization the GRS 1915+105 is mostly originated from the magnetically aligned dust grains within the clouds. Therefore, the jet-origin synchrotron radiation polarization would have given only minor contribution in the NIR band in 2019 Apr-Dec.

## 1. Introduction

GRS 1915+105 (Galactic coordinates: $l = 45°.4$, $b = -0°.2$) is an X-ray binary system consisting of a black hole with a mass of $12.4^{+2.0}_{-1.8} M_\odot$ and a K-type giant companion star with a mass of $0.8 M_\odot$ (Reid et al. 2014; Steeghs et al. 2013). This system is particularly notable for its intermittent relativistic jets (Mirabel & Rodriguez 1994). Polarimetric observations provide a unique insight into the study of jets because high linear polarization is expected from the synchrotron radiation they emit.

Optical polarimetric observations of GRS 1915+105 is too difficult because of the considerable interstellar extinction ($A_V = 19.6 \pm 1.7$ mag; Chapuis & Corbel 2004). On the other hand, the object remains detectable in the near-infrared regime. Recent $K_s$ band polarimetric measurements of GRS 1915+105 conducted from April to December 2019 exhibited stable polarization with $2.42\% \pm 0.08\%$ and position angles of $38°.1 \pm 1°.2$ (Imazato et al. 2021). The detected polarization is probably a synthesis of that of synchrotron radiation from the jet and interstellar polarization produced by the interstellar dust. Estimation of the interstellar polarization component is essential for extracting the jet-origin polarization and examining the jet.

In this study, we performed $K_s$ band polarimetric observations of the field stars in a 10' × 7' field surrounding GRS 1915+105. By utilizing the Gaia DR3 and StarHorse2 catalogs, we analyze the distance dependence of the polarization to estimate the interstellar component toward GRS 1915+105.

This Paper is structured as outlined below. We give the observations and data reduction methodology in Section 2. In Section 3 we present the result and discussion. We summarize the study in Section 4.

## 2. Observations and Data Reduction

Polarimetric observations were performed with the 1.5 m "Kanata" telescope at Higashi-Hiroshima Observatory of Hiroshima University. We used the Hiroshima Optical and Near-InfraRed camera (HONIR; Akitaya et al. 2014) attached to the telescope. Data were collected on 2023 April 10 and October 23 using the $K_s$ band filter.

HONIR's polarimetric optics include a rotatable half-wave plate and a stationary Wollaston prism, facilitating concurrent imaging of ordinary and extraordinary rays on a 2k×2k pixel HgCdTe detector array. A focal mask consisting of 5 rectangular apertures with each 0.75' × 9.8' slots was employed to prevent the overlap of ordinary- and extraordinary-ray images. A series of polarization measurements consisted of four frames captured at half-wave plate position angles of 0.0°, 45.0°, 22.5°, and 67.5°, respectively. Exposure durations were 50 seconds per frame on April 10 (four sets) and 80 seconds per frame on October 23 (three sets), incorporating dithering along the long side of the slots. Because the WCS fitting is difficult for the images taken in polarimetry mode, we took the R band imaging data as well as $K_s$ band one.

In the data reduction, we performed a standard pre-processing procedure of HONIR data, including dark subtraction, flat fielding, and cosmic ray removal to the taken FITS images. Then, aperture photometry was performed for all point sources in the frames. Stokes $I$, $Q/I$, and $U/I$ were computed for each dataset. In this paper, hereafter, $Q/I$ and $U/I$ are simply denoted as $Q$ and $U$, respectively. Multiple data sets derived from dithering observations were utilized to match each star, and the mean and standard deviation of Q and U were calculated. Consequently, individuals with a single measurement point and those exhibiting standard deviations of Q and U exceeding 0.01 (i.e., 1%) were considered unreliable and subsequently excluded. The observation error was determined using standard deviation of Q and U. The degree of polarization (P) and polarization azimuth angle (PA) for each sample were calculated from the mean and standard deviation of Q and U (see Appendix in Kawabata et al. 1999)

Because our polarimetric observations were conducted in the $K_s$ band with the discrete apertures and many stars around GRS 1915+105 suffers from significant interstellar absorption, we performed the astrometry of the stars in the following way. First, we y calculated the WCS using $Astronetry.net$[1] for the R band imaging data, and then determined the WCS for the $K_s$ band imaging data employing the coordinates of the stars present in both the $K_s$ and R band images. Subsequently, we compared the stellar positions in $K_s$ band images and those of the polarimetric images, and derived conversion formula from pixel coordinates to equatorial ones. This method allowed us to acquire equatorial coordinates for each sample star with an error margin of approximately 2 arcseconds across the entire field of view.

We cross-matched the obtained sample stars with the Gaia DR3 catalog and the StarHorse2 catalog. StarHorse2 catalog employs photometric and spectral data sourced externally from Gaia and applies a stellar evolutionary model to more accurately estimate the parameters of each star (Queiroz et al. 2023).

In this paper, the sample stars that were successfully cross-matched with the StarHorse2 catalogue and that are brighter than 19 in G magnitudes were designated as Matched-Data, while the sample stars that could not be cross-matched with the Gaia DR3 catalog are classified as Unmatched-Data.

As a result, 21 samples were categorized as Matched-Data, while 60 samples were designated as Unmatched-Data. Note that all of the Matched-Data stars were located within 4 kpc. The Unmatched-Data stars are not considered to be listed in the Gaia and/or StarHorse2 catalogs due to their significant interstellar absorption in the optical regime; consequently, most Unmatched-Data stars are presumed to be more distant (i.e., beyond 4 kpc) than the Matched-Data sample stars, although some may be nearby stars only bright in NIR bands.

## 3. Result and Discussion
### 3.1 Overview of Observation Results

Figures 1 and 2 show the polarization vector map of the Matched-Data and Unmatched-Data, respectively.

The Matched-Data samples exhibit a low P value of $\leq 0.015$ and a somewhat dispersed PA, whereas the Unmatched-Data samples demonstrate a comparatively high degree of polarization with $P \geq 0.015$ and aligned polarization angle at $PA \cong 40°$.

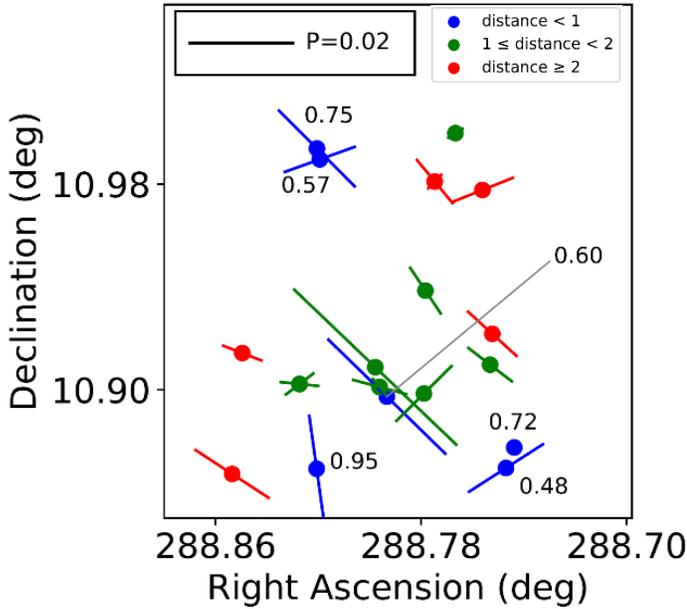

**Fig. 1. Polarization vector map of the Matched-Data.** The horizontal and vertical axes denote the right ascension and declination, respectively. The length of the line for each sample star is proportional to polarization degree, P, while its orientation represents their position angle, PA. Stars located at distances under 1 kpc are depicted in blue, 1-2 kpc in green, and those exceeding 2 kpc in red. The stars of which the distances are less than 1 kpc, the distance (in kpc) are indicated in the panel.

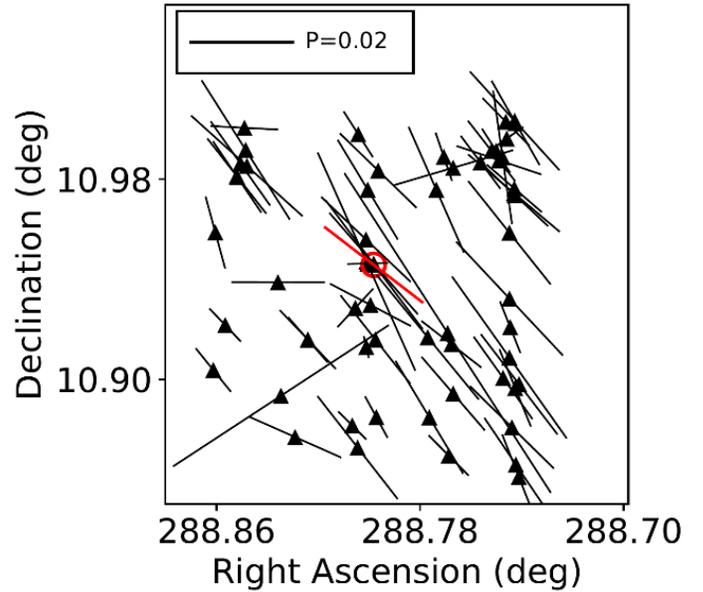

**Fig. 2. Polarization vector map of the Unmatched-Data.** The red open circle shows the mean polarization of GRS 1915+105 (Imazato et al. 2021).

Figure 3 displays the distributions of the polarization parameters for the sample stars. Table 1 presents the mean and standard deviation for each parameter. The results for the PAs in Table 1 indicate that the Matched-Data generally exhibits, on average, lower P values and greater dispersion in PA. Figure 3 demonstrates two peaks at $PA \sim 45°$ and $PA \sim 125°$. More comprehensive statistics indicate 43% of the sample stars fall within the range of $48°.1 \pm 8°.7$ and that 33% of them are within $123°.1 \pm 13°.4$. Conversely, in the Unmatched-Data, the peak around $PA \sim 125°$ is unclear compared to the Matched-Data. 68% of the Unmatched-Data are in the range of $39°.1 \pm 11°.8$.

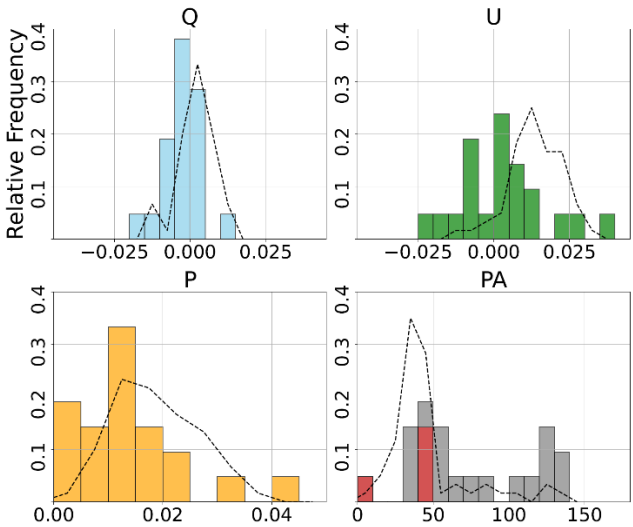

**Fig. 3. Distributions of polarization data.** The upper left, right, and lower left and right panels are for Q, U, P, and PA, respectively. The distributions represented by the bars correspond to the Matched-Data, while those depicted by the dotted lines are for the Unmatched-Data. Concerning PA for the Matched-Data, the samples exhibiting large polarizabilities ($P \geq 0.02$) in the Matched-Data PAs are highlighted in the red bars.

|    | Matched-Data        | Unmatched-Data      |
|----|---------------------|---------------------|
| Q  | $-0.002 \pm 0.006$  | $+0.004 \pm 0.009$  |
| U  | $+0.003 \pm 0.015$  | $+0.014 \pm 0.013$  |
| P  | $0.004 \pm 0.012$   | $0.014 \pm 0.013$   |
| PA | $64°.3 \pm 75°.7$   | $37°.7 \pm 18°.3$   |

**Table 1. Mean and standard deviation of polarization parameters for samples in Matched-Data and Unmatched-Data.**

### 3.2 Observed Distance Dependence of Polarization

Figure 4 illustrates the distance dependence of the attenuation and polarization parameters for each sample star in Matched-Data. The interstellar extinction ($A_G$) increases with distance, exhibiting notable linearity up to 1 kpc. In the polarization parameters, the Q seems to cluster between -0.01 and 0.00 at any distance. Conversely, U demonstrates a scatter, varying from -0.025 to +0.040 for stars, especially within 1.5 kpc (standard deviation $\sigma_U = 0.021$). For stars beyond 1.5 kpc $\sigma_U$ decreases to 0.008. This trend corresponds with P, that keeps around 0.02 within 1 kpc but progressively diminishes with distance, remaining around 0.01 beyond 1.5 kpc.

The results suggest that interstellar clouds generating polarization levels of P=0.01 are distributed within the distance range of 3–4 kpc toward GRS 1915+105. The orientation of the interstellar magnetic field projected onto the sky fluctuates with distance or location, yet the overall distribution exhibits moderate peaks around PA $\cong 45°$ and $\cong 125°$. This elucidates the reduction in the degree of polarization for more distant background stars: despite an increase in the number of interstellar clouds intersecting along the line of sight, resulting in larger extinction, the polarization vectors tend to be canceled our each other. The observed larger dispersion in PA indicates that the interstellar magnetic field in this range are mostly along the line of sight (leading to less tangential component) or lacks a coherent magnetic field structure.

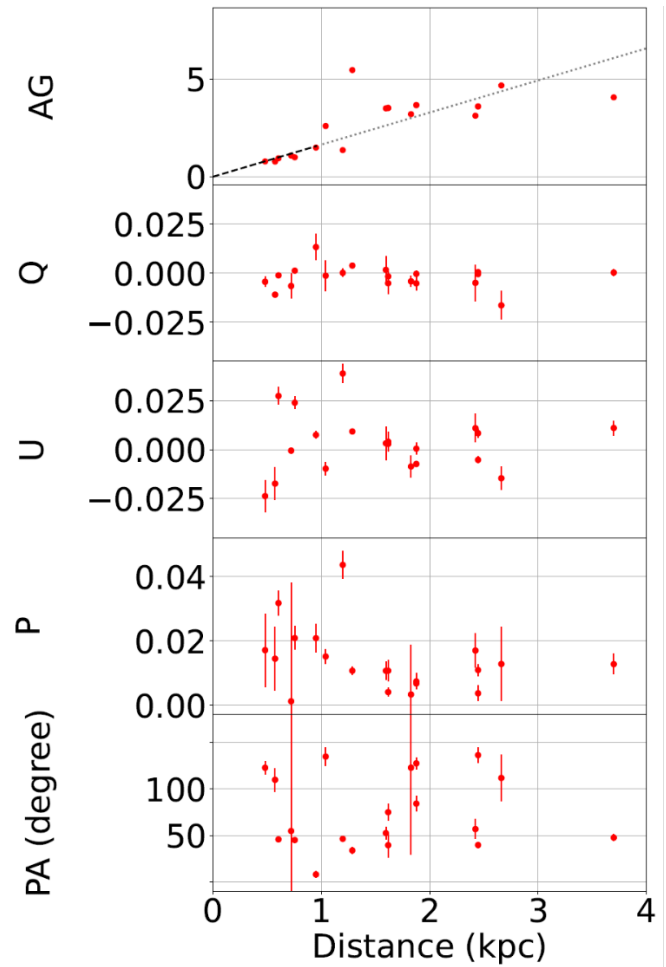

**Fig. 4. Distance Dependence of Polarization and Extinction in Matched-Data.** From top to bottom panels, the distance dependence of interstellar

extinction ($A_G$; from the Gaia catalog), observed Q, U, P and PA are shown, respectively.

Next, we examine the polarization characteristics of the Unmatched-Data. Although we have no direct information on the distance for those samples, as outlined in §2, the Unmatched-Data primarily comprises sample stars located at beyond 4 kpc. Figure 2 demonstrates that the polarization vectors are adequately aligned, and Table 1 reveals a relatively high mean polarization degree ($P = 0.014 \pm 0.013$), accompanied by minimal dispersion in polarization position angle ($PA = 37°.7 \pm 18°.3$). These are consistent with the mean polarization observed for GRS 1915+105 ($P = 0.0242 \pm 0.0008$ and $PA = 38°.1 \pm 1°.2$; Imazato et al. 2021) within the standard deviation, indicating that GRS 1915+105 is analogous to the sample stars in the Unmatched-Data regarding interstellar polarization. These findings suggest that the interstellar clouds responsible for the substantial extinction of these sample stars at $\geq 4$ kpc are penetrated by a coherent magnetic field oriented along $PA \cong 45°$. Thus, the polarization observed in GRS 1915+105 is ascribed to such an interstellar polarization, while the intrinsic polarization resulting from phenomena such as jets is probably less substantial.

### 3.3 Distribution of Interstellar Matter in the Direction of GRS 1915+105

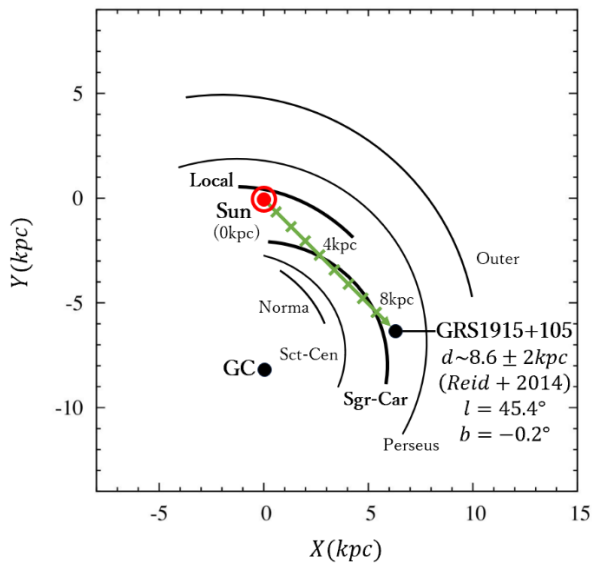

**Fig. 5.** A diagram illustrating the structure of spiral arms in the Milky Way Galaxy and the positional relationship between the Sun and GRS 1915+105. The structure of the spiral arms are referred from VERA collaboration et al. (2020). The red dot represents the position of the sun, and the black dot denotes the possible location of GRS 1915+105. The green arrow from the sun to GRS 1915+105 is ticked at intervals of 1 kpc.

Figure 5 illustrates a schematic representation of the structure of the Milky Way galaxy around the sun and GRS 1915+105. The spiral arm configuration is derived from Hirota et al. (2020). Toward GRS 1915+105, the influence by the interstellar matter associated with the Local Arm (say, < 1kpc) and the inter-arm region (2-4 kpc) are expected. Beyond 4 kpc, the interstellar matter associated with the Sagittarius-Carina Arm is expected. If we assume the distance of $8.6 \pm 2$ kpc for the GRS 1915+105, it would be located within or beyond the Sagittarius-Carina Arm.

Our observations suggest that the interstellar clouds scattered within the inter-arm region up to a distance of 3–4 kpc and the magnetic field component perpendicular to the line of sight are not aligned so much. Besides, optically thick clouds locate within the Sagittarius-Carina Arm and in these clouds the magnetic field component perpendicular to the line of sight is almost uniform at approximately $PA \cong 40°$, which corresponds to $PA \cong 95°$ in galactic coordinates. This orientation is almost parallel to the galactic plane.

In the Matched-Data, the most distant sample locates at a distance of 3.6 kpc, exhibiting an extinction $A_G$ of approximately 4 mag. On the other hand, GRS 1915+105 locates at a distance of $8.6 \pm 2$ kpc with an extinction $A_V = 19.6 \pm 1.7$ mag, indicating a notable increase of the extinction within 3.6-8.6 kpc. This indicates that the stars in the Unmatched-Data exhibit significantly larger extinction than those in the Matched-Data. The line of sight in this direction runs parallel to the Sagittarius-Carina Arm; however, the observation of the well-aligned magnetic field with $PA \cong 45°$ indicates that the interstellar magnetic field in this are possesses a substantial component perpendicular to the arm.

## 4. Summary

We performed NIR polarimetry of the field stars around GRS 1915+105 to estimate the interstellar component of the $K_s$ band polarization that was observed in 2019. We derived the distance-dependence of the interstellar extinction and the polarization parameters of the field stars. The sample stars matched with the Gaia DR3 catalog and the StarHorse2 are within 4 kpc and show the linear increase of the extinction with the distance; however the position angles of them are rather dispersed and the polarization remains smaller, around $P \approx 0.01$ beyond 1.5 kpc. On the other hand, the sample stars that are not matched with the catalogs, that is , the observed samples that are bright enough in $K_s$ band for polarimetry but too faint in optical bands to be observed in Gaia survey, show quite a uniform polarization pattern along $PA = 38° \pm 18°$. Those suggest that the interstellar clouds existing beyond 4 kpc causes the large extinction and that the polarization the GRS 1915+105 is mostly originated from the magnetically aligned dust grains within the clouds. Therefore, the average $K_s$ band polarization degree of $0.0242 \pm 0.0008$ and position angle of $38°.1 \pm 1°.2$ observed during the period from April to December 2019 by Imazato et al. (2021) are primarily attributed to the interstellar polarization induced by interstellar clouds and magnetic fields within the Sagittarius-Carina Arm along the line of sight.

**Acknowledgement**

We thank the support of the observation for Kanata Telescope team, especially for Mr. Taishu Kayanoki, Mr. Ryo Imazawa, Mr. Tomoya Hori and Dr. Tatsuya Nakaoka. This work was supported by the Kagaku Wakuwaku Project (a program promoting sciences for children), the Mazda Foundation.